\documentclass[prb,aps,epsf,twocolumn,showpacs,10pt]{revtex4-1}
\usepackage{graphicx,amsfonts,times,bm,amsmath,verbatim,color,array}

\begin{document}
\title{Yu-Shiba-Rusinov states and topological superconductivity in Ising paired superconductors}

\author{Girish Sharma}
\author{Sumanta Tewari}

\affiliation{Department of Physics and Astronomy, Clemson University, Clemson, SC 29634}

\begin{abstract}
An unusual form of superconductivity, called Ising superconductivity, has recently been uncovered in mono- and few-layered transition metal dichalcogenides. This 2D superconducting state is characterized by the so-called Ising spin-orbit coupling (SOC), which produces strong oppositely oriented effective Zeeman fields perpendicular to the 2D layer in opposite momentum space valleys. We examine the Yu-Shiba-Rusinov (YSR) bound states localized at magnetic impurities in Ising superconductors and show that the unusual SOC manifests itself in unusually strong anisotropy in magnetic field response observable in STM experiments. For a chain of magnetic impurities with moments parallel to the plane of Ising superconductors we show that the low energy YSR band hosts topological superconductivity and Majorana excitations as a direct manifestation of topological effects induced by Ising spin-orbit coupling.
\end{abstract}

\maketitle

\section{Introduction.}
An unusual form of Cooper pairing, called Ising pairing, has recently been uncovered in two-dimensional superconducting states in mono- and few-layered transition metal dichalcogenides (TMDs). TMDs are materials with a 2D honeycomb lattice similar to graphene~\cite{Neto:2009, Sarma:2011}, but with broken in-plane mirror symmetry, resulting in a special type of intrinsic spin-orbit coupling (SOC), called Ising SOC~\cite{{Xiao:2012},
{Zhu:2011}, {Kormanyos:2013}, {Zahid:2013}, {Cappelluti:2013}}. Ising SOC acts as an effective Zeeman field which strongly polarizes the electron spins perpendicular to the 2D plane. This is in stark contrast to the more familiar Rashba SOC, which produces a 2D helical liquid with electron spins polarized in the in-plane directions.  In TMDs the spin polarizations due to Ising SOC are in opposite directions near opposite momentum space valleys ($\mathbf{K}$ and $-\mathbf{K}$), keeping time reversal symmetry intact, unlike in the case of a conventional Zeeman coupling. In this work, we describe unusual effects of SOC on 2D superconductivity in TMDs (called Ising superconductivity~\cite{Xi:2015, Ye:2012, Taniguchi:2012, Shi:2015, {Lu:2015},{Saito:2015},{Xi2:2015}}), including predicting a topological superconducting (TS) phase with Majorana fermion excitations for a chain of magnetic impurities with moments parallel to the 2D plane. Our theoretical predictions, besides being of immediate experimental interest, makes the study of Ising superconductivity important for fundamental physics as well as applications.

Since in TMDs the spin polarizations in opposite momentum space valleys are opposite, Ising SOC favors inter-valley pairing, where electrons with opposite momenta and spin from valleys centered around $\mathbf{K}$ and $-\mathbf{K}$ form Cooper pairs (Ising pairing)~\cite{{Lu:2015},{Saito:2015},{Xi2:2015}, Moratalla:2016, Zhou:2015}. For conventional spin-singlet superconductors it is well known that superconductivity is quenched under the application of a magnetic field. Ignoring orbital effects of the magnetic field, the quenching of superconductivity is due to Zeeman coupling of the magnetic field to electron spins, and can be estimated by equating the binding energy of Cooper pairs with Zeeman splitting (Pauli limit).
In Ising superconductivity the intrinsic SOC protects the electrons from alignment with external magnetic field when it is applied parallel to the plane. This has been experimentally confirmed with recent observations of in-plane upper critical field of more than six times the Pauli paramagnetic limit in superconducting MoS$_2$ and NbSe$_2$ samples~\cite{{Lu:2015},{Saito:2015},{Xi2:2015}}.
In this work we discuss the experimental signatures of the unusual Ising SOC on magnetic impurity induced Yu-Shiba-Rusinov (YSR) bound states and the emergence of topological superconductivity and Majorana fermion excitations in YSR bands in two-dimensional Ising superconductors.


Magnetic impurities in superconductors can support sub-gap bound states known as Yu-Shiba-Rusinov (YSR) states~\cite{Yu:1965, Shiba:1968, Rusinov:1969}. The mid-gap YSR bound states emerging due to a single localized magnetic impurity located on a $s$-wave superconductor can give rise to a zero bias peak (ZBP) in the local density of states (LDOS) measurement, signaling a level crossing and change in ground state parity of the many-body wave function~\cite{{Balatsky:2006}}. This transition occurs when the impurity strength $J$ is tuned near a critical value $J_c$. The ZBPs induced by YSR states can be split by an external Zeeman field which couples to the spin ~\cite{{Balatsky:2006}}. Here we show that the ZBPs arising from YSR states localized at magnetic impurities
in Ising superconductors are robust to an unusually high in-plane magnetic field. We demonstrate with a $T$-matrix calculation that Ising SOC is directly responsible for the anomalously large critical in-plane magnetic field where the ZBP splits away from zero energy. This behavior, which correlates with the anomalously large anisotropy in upper critical fields between magnetic fields applied {perpendicular and parallel to the 2D plane}~\cite{{Lu:2015},{Saito:2015},{Xi2:2015}}, can be tested in STM experiments.
Moreover, for a chain of a dilute concentration of magnetic impurities with moments \textit{parallel} to the plane of Ising superconductors, we establish the emergence of a topological superconducting phase with end-state Majorana fermions by numerical diagonalization of the Bogoliubov de-Gennes (BdG) equations. 
In the complementary band (or `wire') limit, where the impurity orbitals of neighboring adatoms strongly overlap, the impurity chain realizes a ferromagnetic wire with moments parallel to the plane of the superconductor. Since Ising SOC engenders a triplet pair potential with Cooper pair spins parallel to the plane \cite{Zhou:2015}, in this case the ferromagnetic wire becomes a topological superconductor (in BDI class) by proximity effect \cite{Hui:2015,Dumitrescu:2015}. This is similar to the case of a half-metal on Ising superconductor discussed elsewhere \cite{Zhou:2015}. We thus establish Ising superconductors with magnetic adatoms with moments parallel to the plane
as a robust platform for topological phenomena and Majorana fermions. That moments need to be parallel to the plane is a direct consequence of Ising SOC (which is perpendicular to the plane), in marked contrast to superconductors with Rashba SOC where adatom moments need to be {perpendicular to the plane to support topological phases and Majorana fermions}.

This paper is organized as follows: In Section II we introduce the Hamiltonian for MoS$_2$ as a prototype for TMD systems and examine its Fermi surface. Even though we start with a Hamiltonian (Eq. 2) that has only a spin-singlet $s$-wave order parameter, because of the spin-orbit coupling a spin-triplet $p$-wave term is generated in the Green's function of the superconductor (Eq. 6) ~\cite{Gorkov:2001, Frigeri:2004, Brydon:2015, Zhou:2015}. 
In Section III we compute the LDOS for a localized Yu-Shiba-Rusinov state, and study its response to an external magnetic field. In Section IV we discuss how a YSR band, formed with a dilute magnetic impurity chain, can host topological superconductivity. Using the BdG equations for a chain of magnetic atoms embedded in a host Ising superconductor, we explicitly demonstrate the existence of Majorana fermion excitations by exact numerical diagonalization and also by mapping on the one dimensional Kitaev model. We end with discussions and conclusion in Section V. 
\section{Hamiltonian}
\begin{figure}[t]
\includegraphics[scale=.20]{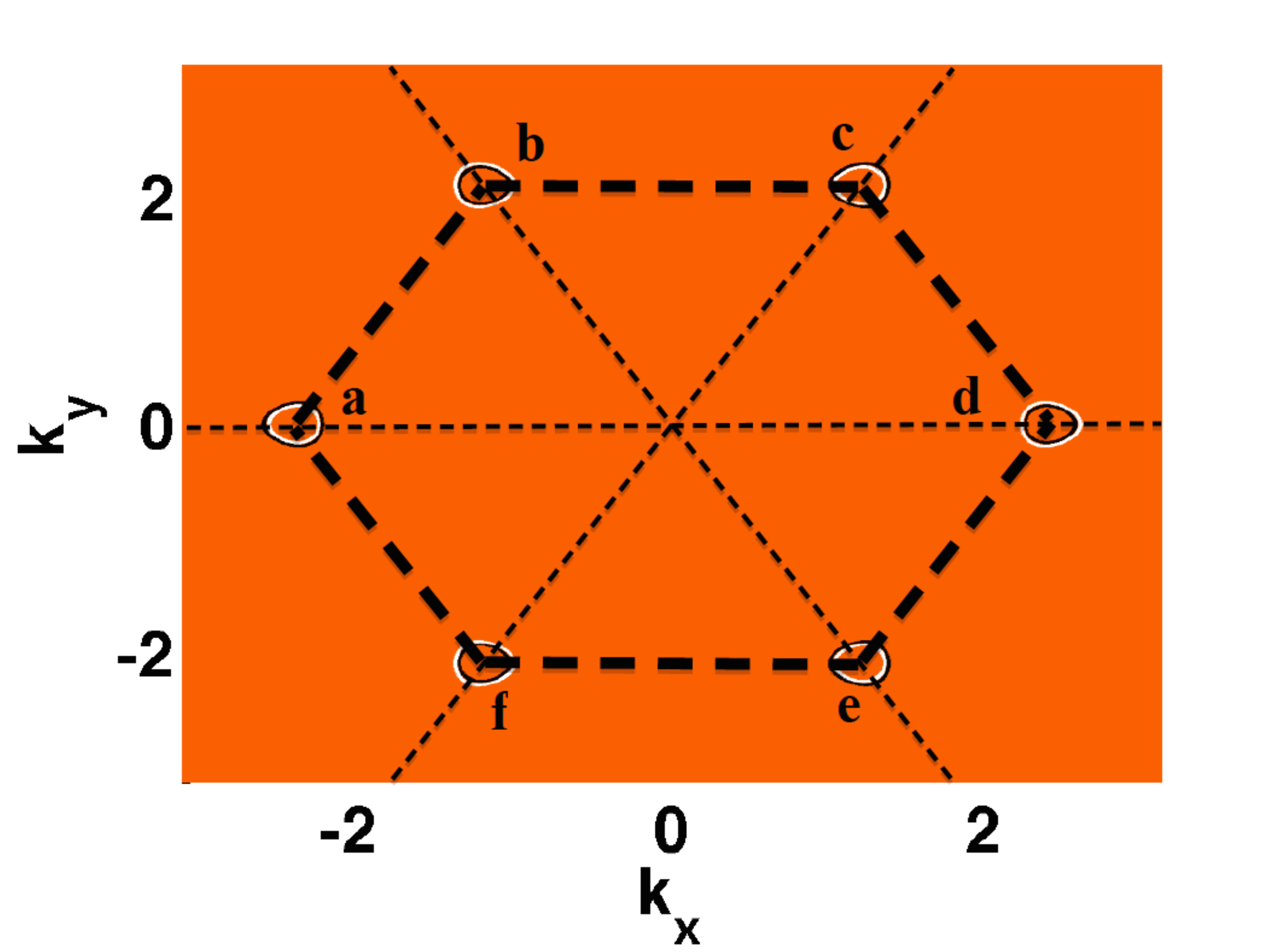}
\includegraphics[scale=.1]{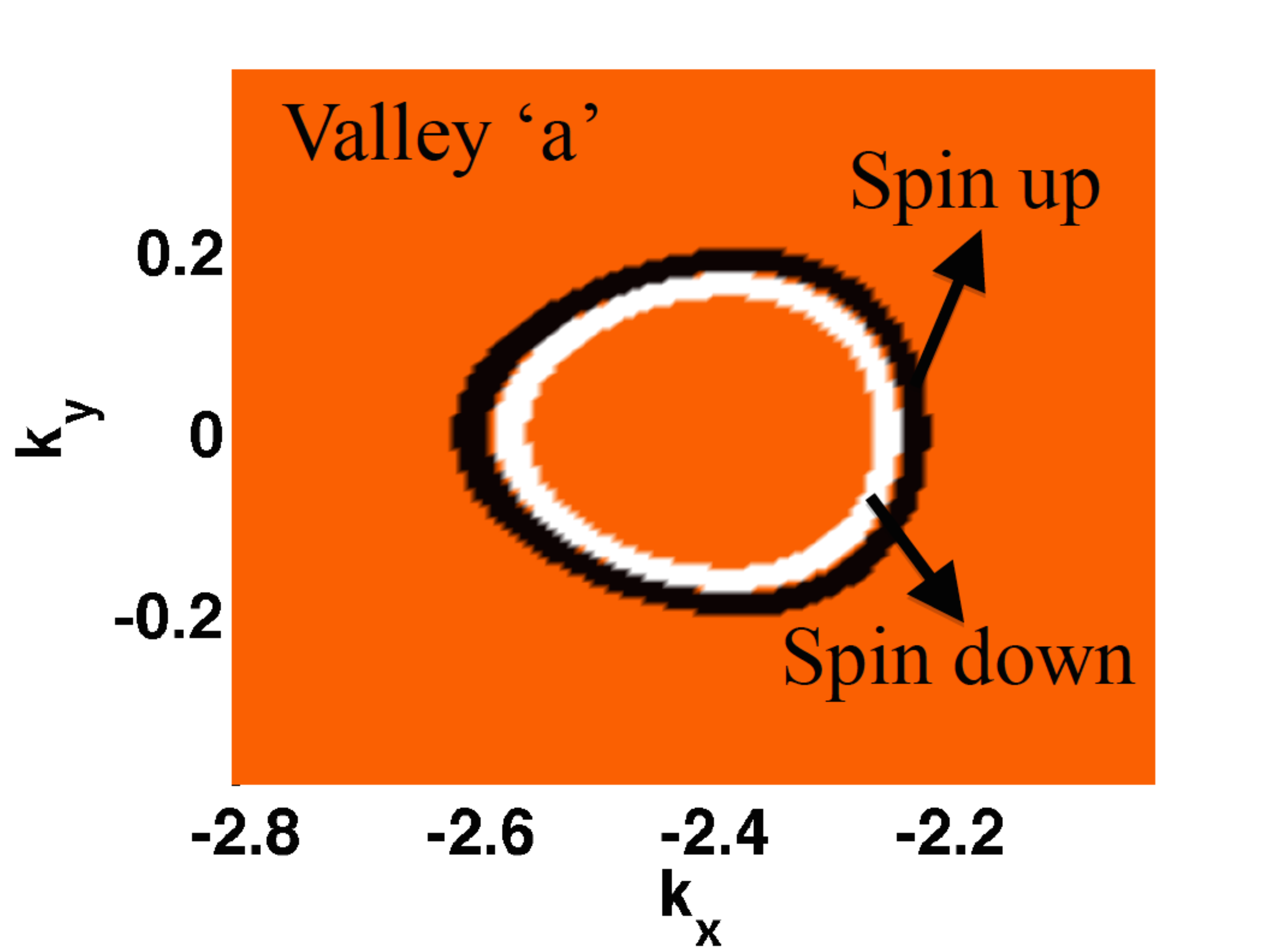}
\includegraphics[scale=.1]{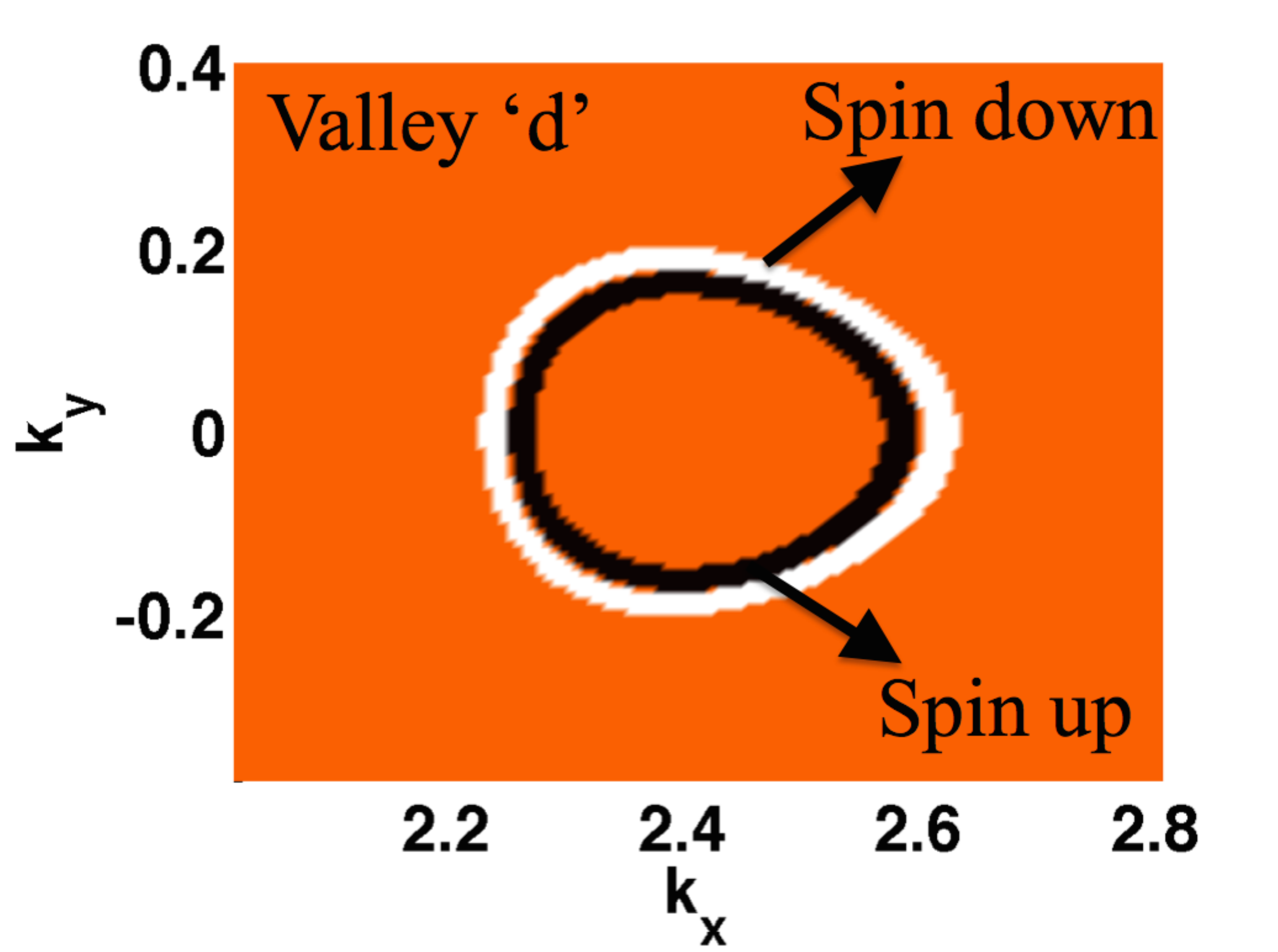}
\caption{\textit{Top panel:} Fermi surface of MoS$_2$ at $\mu=0.25t$ in the presence of Ising SOC with strength $\alpha = 8meV$. The Ising SOC, which acts like an effective Zeeman field, generates spin-polarized Fermi pockets (`white: spin up', `black: spin down') in the vicinity of six valley points ($\mathbf{K}$ points) labeled from `a' to `f'. Since the SOC strength changes sign at each valley, the spin-splitting is also opposite near each valley point. \textit{Bottom panel:} Fermi surfaces near two valley points `a' and `d' have been zoomed to clearly illustrate the opposite spin-splitting occurring near these two points.}
\label{Figure_Fermi_surface}
\end{figure}

We start with the Hamiltonian for a representative Ising superconductor, MoS$_2$, which in the basis $\Psi^\dagger_{\mathbf{k}}=(c^\dagger_{\mathbf{k},\uparrow}, c^\dagger_{\mathbf{k},\downarrow})^T$ can be written as~\cite{Lu:2015, Zhou:2015} $H=\sum\limits_{\mathbf{k}}{\Psi^{\dagger}_{\mathbf{k}} H_0({\mathbf{k}})\Psi_{\mathbf{k}}}$, where
\begin{eqnarray}
H_0({\mathbf{k}}) = \zeta_{\mathbf{k}}\sigma_0 + F_{\mathbf{k}}\sigma_z
\label{Eq_Hk_1}
\end{eqnarray}
The Pauli matrix $\sigma_z$ acts in the spin space and $\sigma_0\equiv I_2$. The operator $c_{\mathbf{k},\sigma}$ annihilates an electron with spin $\sigma$ and momentum $\mathbf{k}$. The function $\zeta_{\mathbf{k}} = t |2\cos(k_x\sqrt{3}/2) e^{ik_y/2} + e^{-3ik_y/2}|-\mu$, is the non-interacting dispersion for MoS$_2$ which generates six valley points in the first Brillouin zone~\cite{Neto:2009} ($\mathbf{K}$ points), where $\zeta_{\mathbf{K}}-\mu=0$. The function $F_{\mathbf{k}} = \alpha (\sin(k_x) - 2\cos(\sqrt{3)}k_y/2)\sin(k_x/2))$ is the Ising SOC term~\cite{Lu:2015, Xi2:2015, Zhou:2015}. Importantly, $F_{-\mathbf{k}} = -F_{\mathbf{k}}$, and therefore the system lacks inversion symmetry.
This particular form of the Ising SOC term $F_{\mathbf{k}}$ suffices to discuss the low energy physics in the vicinity of each valley point, and reproduce the Fermi surface of MoS$_2$.
The nearest neighbor hopping integral $t$ is fixed to $t=0.5 eV$ in this paper. For all our calculations we will choose the SOC strength $\alpha=8 meV$, and $\mu=0.25t$~\cite{Lu:2015}. Figure~\ref{Figure_Fermi_surface} shows the Fermi surface for MoS$_2$ as obtained from Eq.~\ref{Eq_Hk_1}. The spin degeneracy is lifted by the Ising SOC, producing spin-polarized Fermi pockets. Since the SOC strength changes sign near each valley, the spin splitting is also opposite at each valley point. At valley points $+\mathbf{K}$ and $-\mathbf{K}$, the electrons are subjected to an effective Zeeman field in opposite directions ($+\alpha F_{\mathbf{K}}$ and $-\alpha F_{\mathbf{K}}$). One can approximate the low-energy Hamiltonians near  $\mathbf{K}$ and $-\mathbf{K}$ as $H_{\mathbf{k}=\mathbf{k}'+\epsilon \mathbf{K}} \approx \mathbf{k}'^2/2m + \epsilon \alpha |F_{\mathbf{K}}|$, where $\epsilon=+1$/$-1$, for $+\mathbf{K}$/$-\mathbf{K}$ respectively. However, in this paper we will consider the full band structure given by Eq.~\ref{Eq_Hk_1} for all our calculations.

We can now write down the mean-field superconducting Hamiltonian in the presence of a spin-singlet $s-$wave superconducting order parameter ($\Delta$) as:
\begin{eqnarray}
H_{\mathbf{k}} = \left( \begin{array}{cc}
H_0(\mathbf{k}) & \Delta\sigma_0 \\
\Delta\sigma_0 & -\sigma_y H_0^*(-\mathbf{k})\sigma_y
\end{array} \right)
\label{Eq_Hk_2}
\end{eqnarray}
Eq.~\ref{Eq_Hk_2} is written in the Nambu basis $\Psi^\dagger_{\mathbf{k}}=(c^\dagger_{\mathbf{k},\uparrow}, c^\dagger_{\mathbf{k},\downarrow}, c_{-\mathbf{k},\downarrow}, -c_{-\mathbf{k},\uparrow})^T$. For our calculations, we will use $\Delta(T=0)\sim 1.7 k_BT_c$, for a fixed $T_c\sim 10K$ throughout in this paper~\cite{Lu:2015}. 

We now introduce a single localized magnetic impurity with a spin $\mathbf{S}$. Further, we assume the impurity to be purely classical. The impurity Hamiltonian, which describes the interaction between the conduction electrons and the localized magnetic moment, can be written as~\cite{Balatsky:2006}
\begin{eqnarray}
H_{\text{imp}} = -J \mathbf{S}\cdot (\Psi^\dagger (\mathbf{r}=\mathbf{R}) \tau_0\hat{\sigma}\Psi(\mathbf{r}=\mathbf{R})),
\label{Eq_H_imp}
\end{eqnarray}
where $\Psi(\mathbf{r})$ is the Fourier transform of $\Psi_{\mathbf{k}}$, $\mathbf{R}$ is the location of the impurity, and $J$ is the exchange strength. The Pauli matrices $\sigma$ and $\tau$ act in spin and particle-hole space respectively. For simplicity, we will set $\mathbf{R}=0$.

\section{YSR states and anomalous magnetic field response} 
\begin{figure}[t]
\includegraphics[scale=.21]{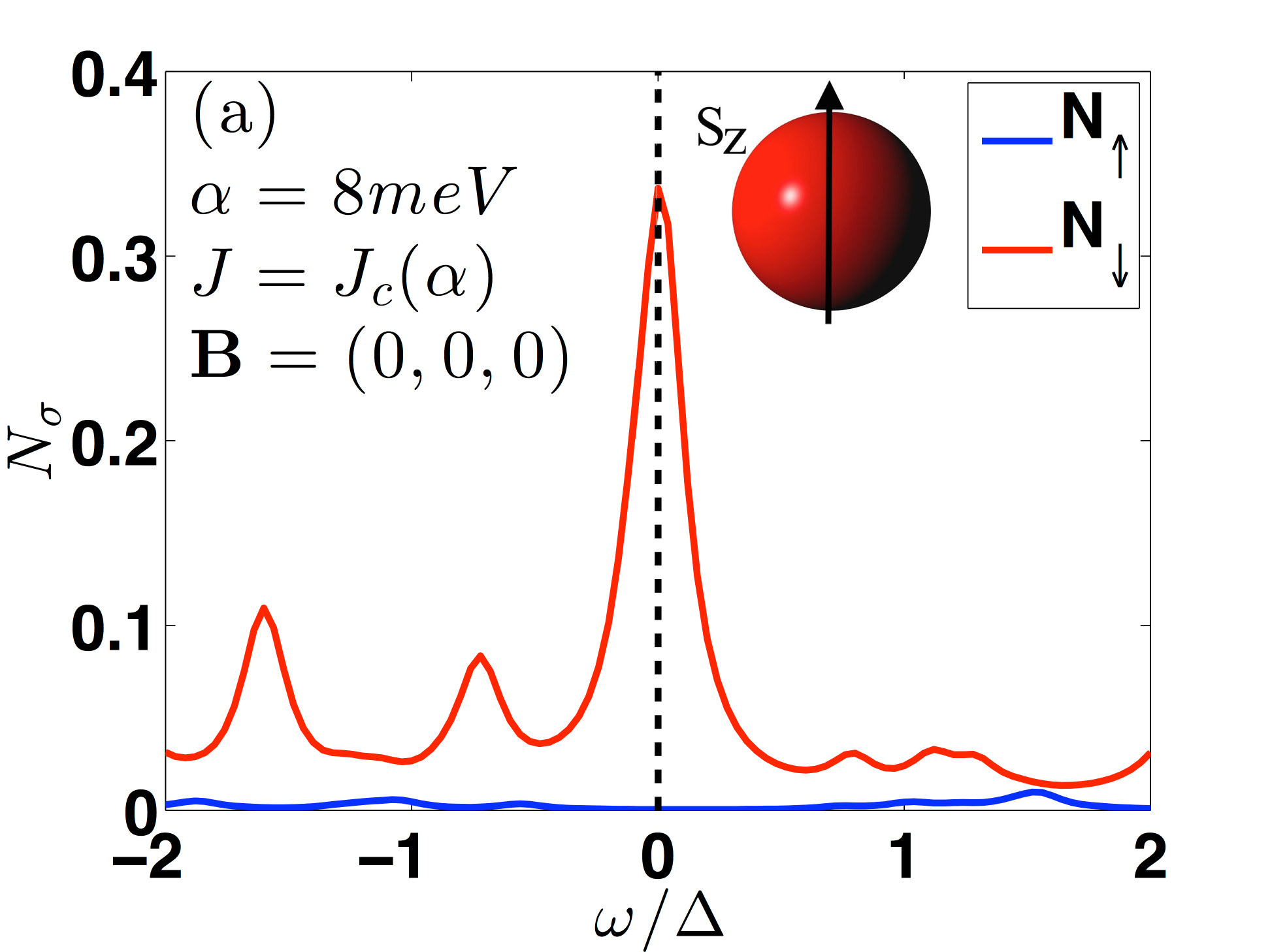}
\includegraphics[scale=.21]{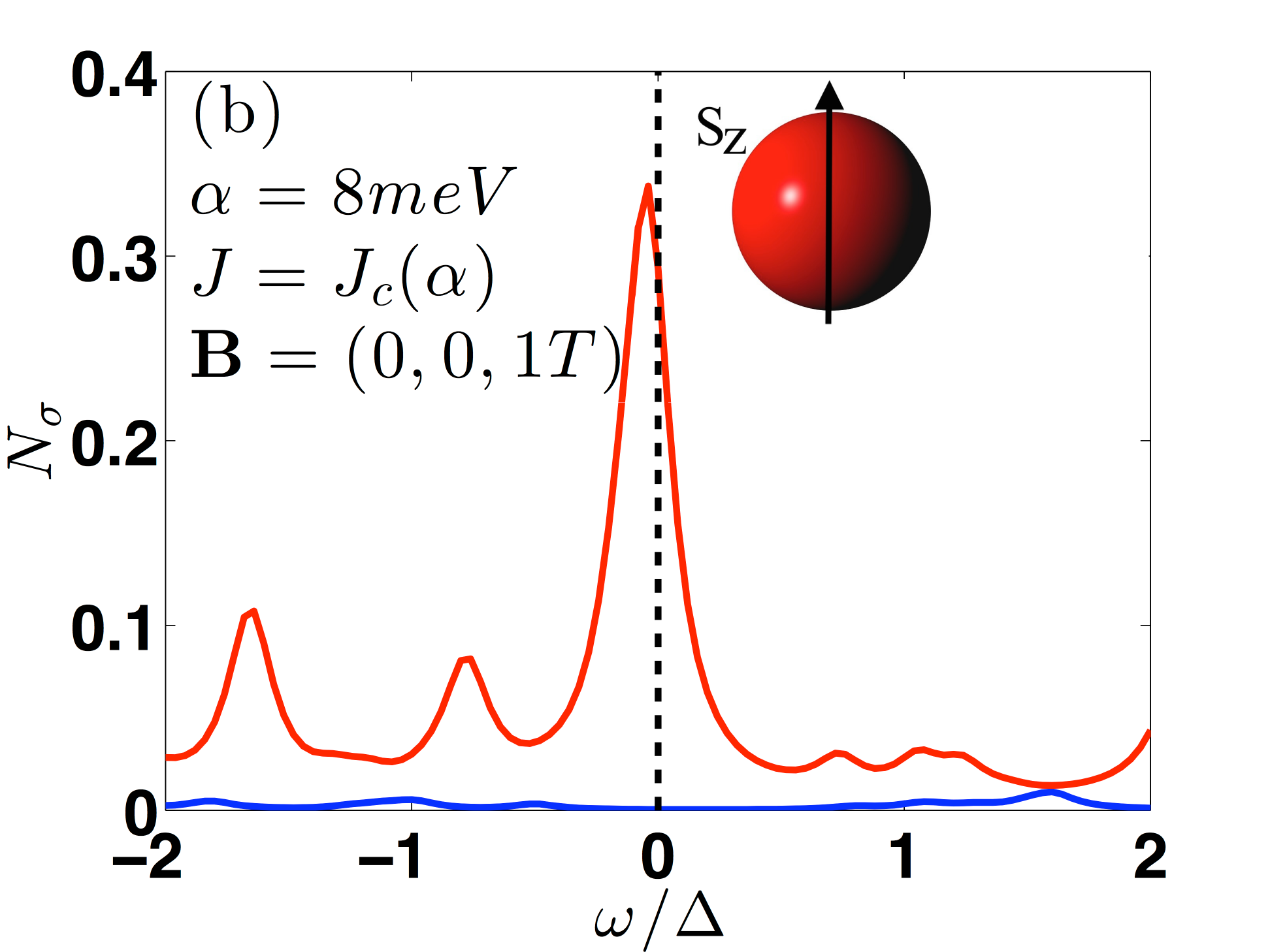}
\includegraphics[scale=.21]{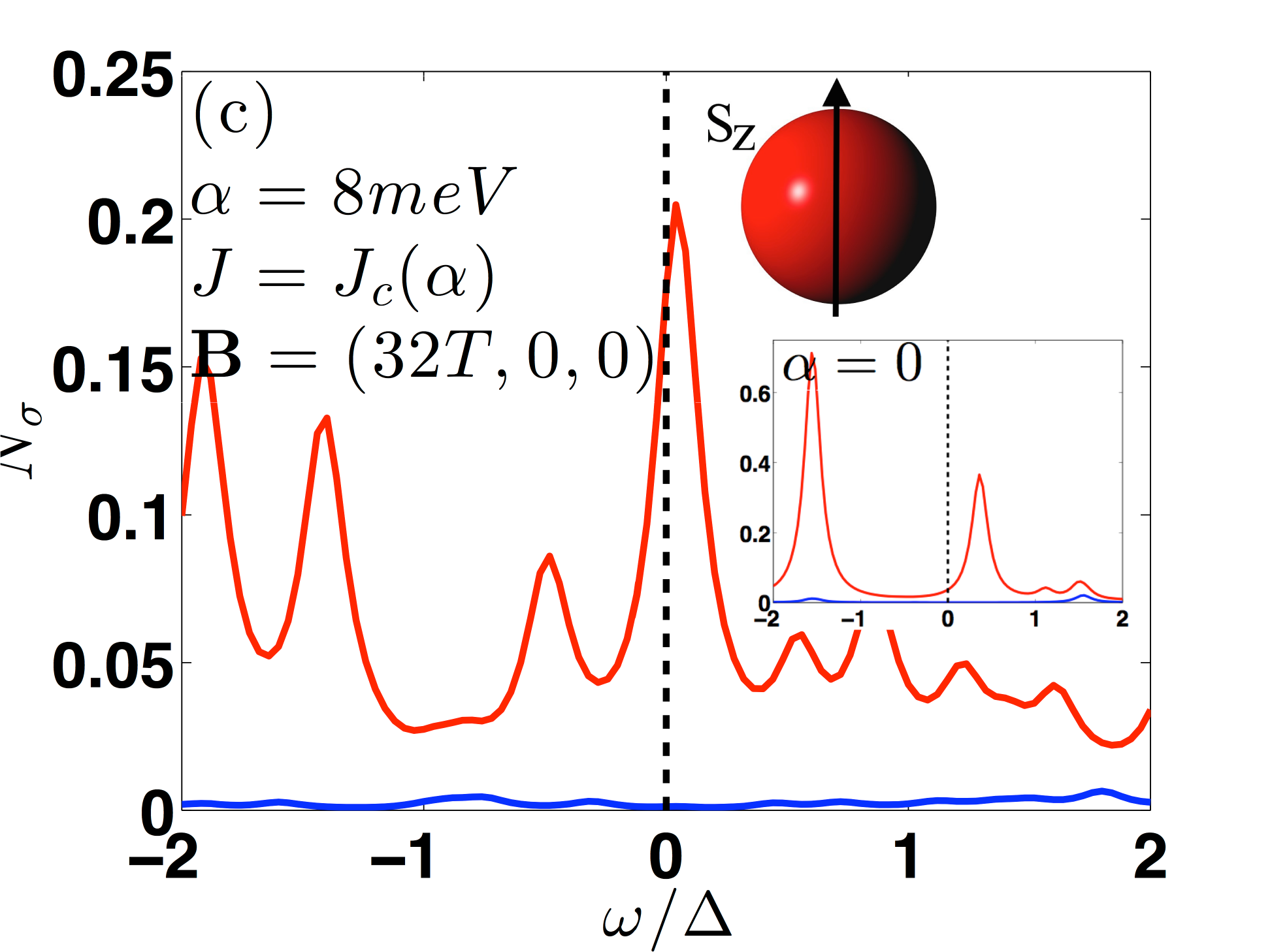}
\includegraphics[scale=.21]{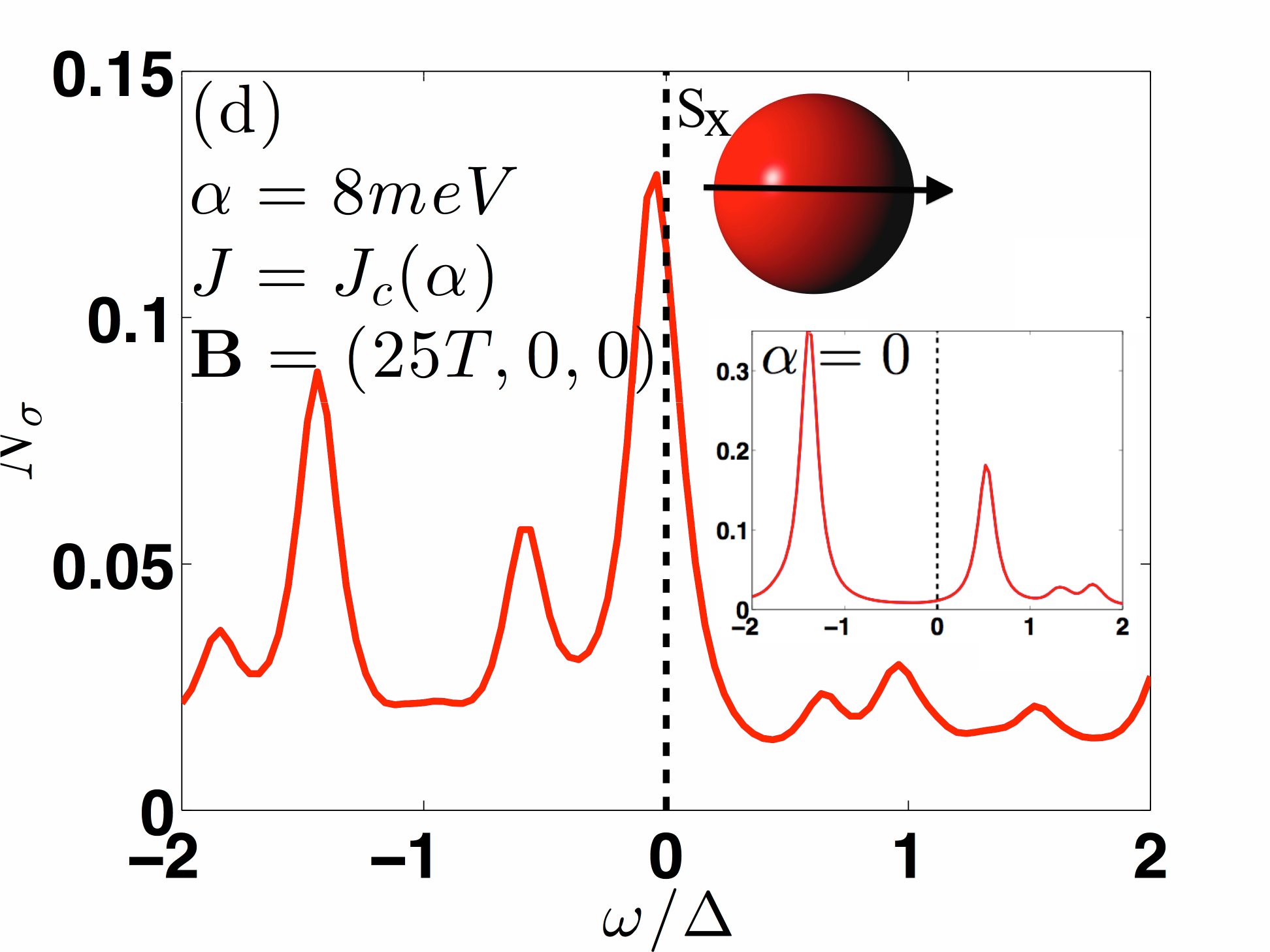}
\caption{(color online). Spin-resolved LDOS in arbitrary units, for a system with magnetic impurity located at $\mathbf{r}=0$. For plots (a)-(c), the impurity spin is pointing in the $z$ direction, for plot (d), the impurity spin points in the $x$ direction. (a) ZBP occurring at a critical value of impurity strength $J_c=J_c(\alpha)$ for both spin-up and spin-down components, with $\alpha=8meV$. (b) A perpendicular magnetic field as small as 1T begins to split the ZBP. (c) The in-plane critical magnetic field has increased to $32T$, when $\alpha=8meV$. The inset shows split ZBP at the same field when $\alpha=0$. (d) The in-plane critical magnetic field (for an impurity spin in $x$ direction) has increased to $25T$, when $\alpha=8meV$. Then inset shows split ZBP at the same field when $\alpha=0$. We have defined the critical field as the value of magnetic field where the ZBP splits from $\omega=0$ to $\omega=\pm 0.05\Delta$. }
\label{Figure_Shiba_states}
\end{figure}

Magnetic moments have pair-breaking effects on a superconducting system, and as a result localized sub-gap excitations emerge~\cite{Yu:1965, Shiba:1968, Rusinov:1969}.
This section is devoted to studying properties of the Shiba state with a single magnetic impurity on superconducting MoS$_2$ surface.
The $T$-matrix approximation is employed to compute the local density of states (LDOS) of YSR states bound at the impurity site~\cite{{Balatsky:2006}, Mahan:2000}. In momentum space, the impurity potential can be written as: $H_{\text{imp}}=\sum\limits_{\mathbf{k},\mathbf{k}'} \Psi^\dagger_{\mathbf{k}'}V_{\mathbf{k},\mathbf{k}'}\Psi_{\mathbf{k}}$, where $V_{\mathbf{k},\mathbf{k}'}$ is the scattering potential: $V_{\mathbf{k},\mathbf{k}'}=-J\mathbf{S}\tau_0\hat{\sigma}$. The $T$-matrix is the solution of the following equation~\cite{{Balatsky:2006}}:
\begin{eqnarray}
T(\mathbf{k},\mathbf{k}',\omega) = V_{\mathbf{k},\mathbf{k}'} + \sum_{\mathbf{k}''}V_{\mathbf{k},\mathbf{k}''}G_0(\mathbf{k}'',\omega) T(\mathbf{k}'',\mathbf{k}',\omega)
\label{Eq_T_matrix_1}
\end{eqnarray}
In Eq.~\ref{Eq_T_matrix_1}, $G_0(\mathbf{k},\omega)$ is the Green's function for the clean system without the magnetic impurity. Once the $T$-matrix is obtained, the Green's function in the presence of the impurity is then given by~\cite{Balatsky:2006}
\begin{eqnarray}
G(\mathbf{r},\mathbf{r}',\omega) = G_0(\mathbf{0},\omega) + G_0(\mathbf{r},\omega)T(\omega)G_0(-\mathbf{r}',\omega),
\end{eqnarray}
where $G_0(\mathbf{r},\omega) = \sum\limits_{\mathbf{k}}G_0(\mathbf{k},\omega)e^{i\mathbf{k}\cdot\mathbf{r}}$. The spin-resolved LDOS can be computed as $N_{\sigma,\mathbf{r}} = -\frac{1}{\pi}{\text{Im }G_{\sigma,\sigma}(\mathbf{r},\mathbf{r},\omega)}$. Figure~\ref{Figure_Shiba_states}a shows the zero bias peak (ZBP) in the density of states of the YSR state for a magnetic impurity with moment perpendicular to the plane, occurring at a particular value of the impurity exchange strength $J_c$. We find similar ZBPs also for magnetic impurities with moments parallel to the plane. Though $J_c$ depends on the material parameters, we specify that $J_c=J_c(\alpha)$, where $\alpha$ measures the strength of the Ising SOC. Therefore in Figure~\ref{Figure_Shiba_states}, the impurity strength has to be tuned to $J_c(\alpha)$ for different values of the SOC parameter to obtain a ZBP.

The effect of an external Zeeman field on these ZBPs can now be studied.  Mathematically, the effect of an external Zeeman field can be introduced by adding the term $H_Z = \mathbf{h}\cdot\boldsymbol{\sigma}\tau_0$ to the Hamiltonian $H_0(\mathbf{k})$ in Eq.~\ref{Eq_Hk_1}. We will assume that the superconducting pairing gap remains unchanged on application of external magnetic field. However relaxing this assumption does not change our results qualitatively. First, we fix the impurity spin $\mathbf{S} = |\mathbf{S}|\hat{z}$, and therefore the Shiba bound states are also spin-polarized along the $z$ direction.  
We find the effects of an applied Zeeman field on the impurity induced ZBPs to be highly anisotropic in the presence of Ising SOC (see Fig.~\ref{Figure_Shiba_states}). For instance when the impurity spin points in the $z$ direction, as shown in Fig.~\ref{Figure_Shiba_states}a, a ZBP appears for a critical impurity strength $J=J_c(\alpha)$, where $\alpha$ is the Ising SOC strength. As shown in Fig.~\ref{Figure_Shiba_states}b, an applied magnetic field parallel to the impurity spin, splits the impurity induced ZBP for a magnetic field strength as low as $\sim 1$ $T$. However as shown in Fig.~\ref{Figure_Shiba_states}c, when the applied field is parallel to the plane of the superconductor (perpendicular to the impurity spin), the magnetic field required to split the ZBP is as high as $\sim 32$ $T$.  In the inset of Fig.~\ref{Figure_Shiba_states}c, we also show the extent of ZBP splitting for $B_x=32$ $T$ and $\alpha=0$. This dramatic enhancement of the anisotropy between the effects of the magnetic field when it is applied parallel and perpendicular to the SOC correlates well with the similar anisotropy in upper critical magnetic field seen in the recent experiments~\cite{{Lu:2015},{Saito:2015},{Xi2:2015}}. 

Some amount of anisotropy in the magnetic field response of an impurity induced ZBP is expected even without spin-orbit coupling. This can be qualitatively explained form a perturbative argument. When the applied magnetic field is parallel to the impurity spin, first order corrections due to the applied field to the energies of the YSR states are  finite, and the critical magnetic field for ZBP splitting is small. However, when the field is perpendicular to the impurity spin, given that the YSR states are polarized in the direction of the impurity spin (in the limit of zero field), the first order corrections to the YSR state energies vanish. In this case a larger applied field is necessary for splitting of the ZBP due to second order effects. We see this anisotropy of magnetic field response of the YSR states even for $\alpha=0$ by obtaining different critical fields (with ratio $\sim 1:8$), for ZBP splitting for the field directions parallel and perpendicular to the impurity spin.  In the presence of non-zero $\alpha$ ($\sim 8meV$), the spins are strongly polarized perpendicular to the $x-y$ plane and the anisotropy of the magnetic field response of the YSR states, as revealed by critical fields of the ZBP splitting, dramatically enhances as shown in Fig.~\ref{Figure_Shiba_states}. This behavior is consistent with a similar effect discussed in references [\onlinecite{{Lu:2015},{Saito:2015},{Xi2:2015}}] for upper critical magnetic fields of the superconducting states.

\section{Topological superconductivity in sub-gap YSR band}
Motivated by recent experiments on topological superconductivity on magnetic impurity chains embedded on a superconductor~\cite{  {Perge2:2014}, Ruby:2015}, we wish to examine the possibility of topological phenomena in dilute chain of magnetic impurities deposited on Ising superconductors. First we will discuss the case of a single magnetic impurity, and then extend our discussion to an 1D array of magnetic impurities arranged in a chain-like fashion. 
We begin with writing the momentum space Green's function for the Hamiltonian in Eq.~\ref{Eq_Hk_2}, which can be expressed as
\begin{eqnarray}
G_0(\mathbf{k},\omega)=\left( \begin{array}{cccc}
\frac{\omega+\zeta_{\mathbf{k}}+F_{\mathbf{k}}}{A^+} &0 & \frac{\Delta}{A^+} & 0 \\
0 & \frac{\omega+\zeta_{\mathbf{k}}-F_{\mathbf{k}}}{A^-} & 0 & \frac{\Delta}{A^-} \\
\frac{\Delta}{A^+} & 0 & \frac{\omega-\zeta_{\mathbf{k}}-F_{\mathbf{k}}}{A^+}  & 0\\
0 & \frac{\Delta}{A^-}  & 0 & \frac{\omega-\zeta_{\mathbf{k}}+F_{\mathbf{k}}}{A^-}
\end{array} \right) \nonumber\\
\label{Eq_Green_fn_1}
\end{eqnarray}
where 
\begin{eqnarray}
A^{\pm} = \omega^2 - \zeta_{\mathbf{k}}^2 - F_{\mathbf{k}}^2 \mp2F_{\mathbf{k}}\zeta_{\mathbf{k}}-\Delta^2
\label{Eq_A_pm}
\end{eqnarray} 
In obtaining Eq.~\ref{Eq_Green_fn_1} we have used the fact that the Ising SOC function $F_{\mathbf{k}}$ is inversion asymmetric. As a result of Ising SOC, the Green's function contains a mixture of both singlet and triplet terms~\cite{Zhou:2015, Gorkov:2001, Frigeri:2004} in the superconducting order parameter, as $A^+\neq A^-$ when $F(\mathbf{k})\neq 0$.  The spin-triplet pairing correlation is given by $\Delta_T(\mathbf{k},\omega) = 4\Delta F_{\mathbf{k}}\zeta_{\mathbf{k}} /A^+A^-$. The $d$-vector, which parametrizes the spin-triplet pairing is parallel to the direction of Ising SOC.  

The Bogoliubov-de Gennes equation for the superconducting Hamiltonian (Eq.~\ref{Eq_Hk_2}) in the presence of a single  localized impurity potential (Eq.~\ref{Eq_H_imp}) is: $(H+H_{\text{imp}})\Psi(\mathbf{r}) = \omega\Psi(\mathbf{r})$. We will be interested in impurity states which are deep in the gap: $|\omega|\ll\Delta$. The BdG equation can be rewritten in the following form~\cite{Pientka:2013, Brydon:2015}
\begin{eqnarray}
(\mathbb{I} + G_0(\mathbf{r}=\mathbf{0},\omega) J\mathbf{S}\tau_0\hat{\sigma})\Psi(\mathbf{\mathbf{r}=0}) = 0,
\label{Eq_bdg_1}
\end{eqnarray}
\begin{figure}
\includegraphics[scale=.3]{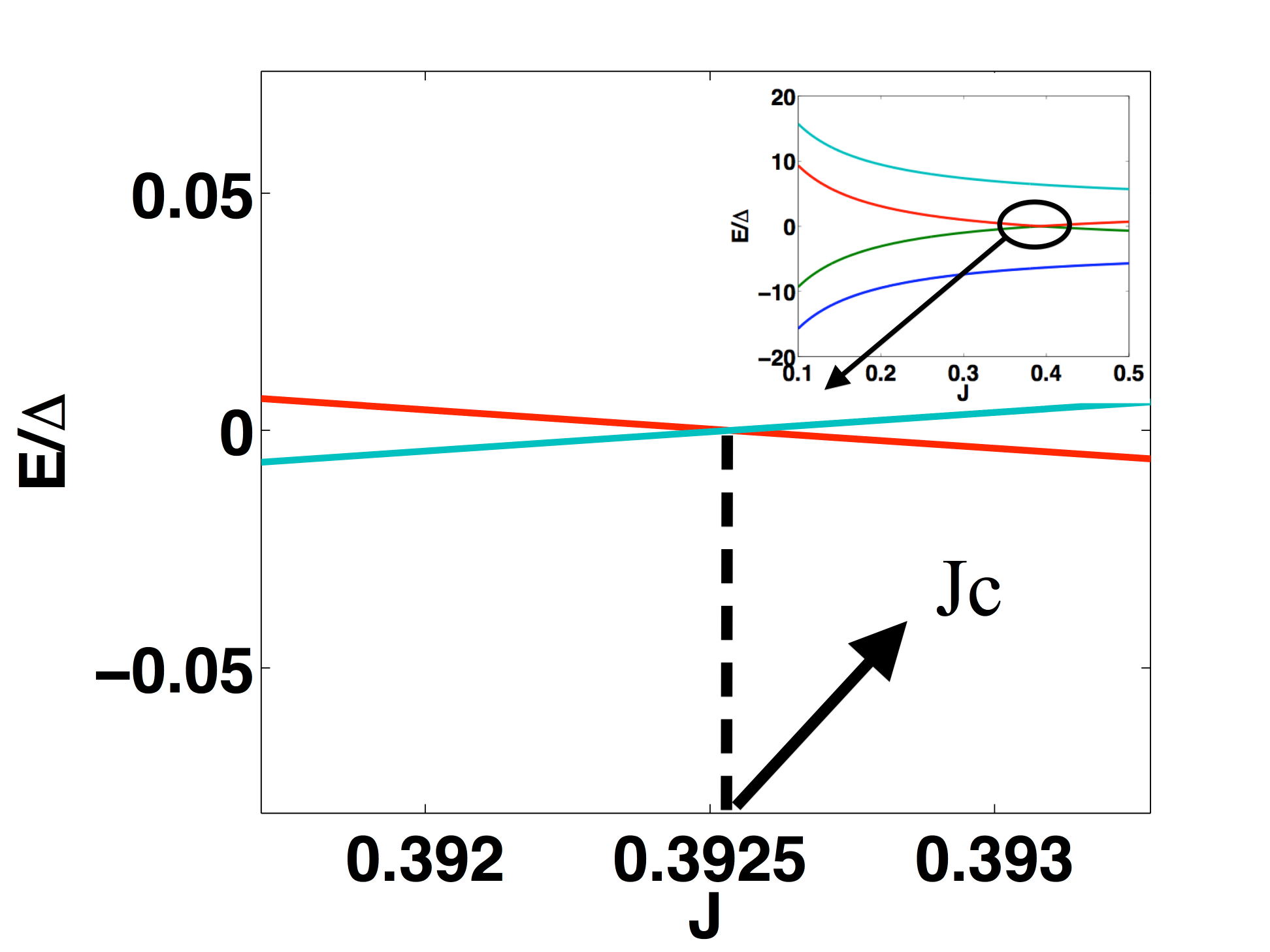}
\caption{(color online). Energy spectrum for a single localized impurity YSR state as a function of the impurity strength $J$, as obtained by numerical solution of Eq.~\ref{Eq_Heff_1}. The inset shows the four energy levels for a wide range of $J$. The main figure illustrates the two low-energy levels which cross each other at zero energy at a certain critical value of impurity strength $J_c$. The spin of the impurity is assumed to be aligned along the $x$-direction.}
\label{Fig_Shiba_spec}
\end{figure}
where $G_0(\mathbf{r},\omega)$ is the real space Green's function obtained by Fourier transforming Eq.~\ref{Eq_Green_fn_1}. 
We need to evaluate the following integrals in order to calculate $G_0(\mathbf{r}=0,\omega)$ from Eq.~\ref{Eq_Green_fn_1}
\begin{eqnarray}
&J_0^{\pm} = \int{\frac{[d^2\mathbf{k}]}{A^{\pm}}}; I_0^{\pm} = \int{\frac{[d^2\mathbf{k}]\zeta_{\mathbf{k}}}{A^{\pm}}};
&K_0^{\pm} = \int{\frac{[d^2\mathbf{k}]F_{\mathbf{k}}}{A^{\pm}}}
\end{eqnarray}
Using the above definitions and Eq.~\ref{Eq_Green_fn_1}, the Green's function $G_0(\mathbf{r}=0,\omega)$ takes the form $G_0(\mathbf{0},\omega)=$
\tiny
\begin{eqnarray}
\left( \begin{array}{cccc}
\omega J_0^+ + I_0^+ + K_0^+ &0 & \Delta J_0^+ & 0 \\
0 & \omega J_0^- + I_0^- - K_0^-  & 0 & \Delta J_0^-\\
\Delta J_0^+ & 0 & \omega J_0^+ - I_0^+ - K_0^+  & 0\\
0 &\Delta J_0^-  & 0 & \omega J_0^- - I_0^- + K_0^-
\end{array} \right) \nonumber\\
\end{eqnarray}
\normalsize
From the functional forms of $F_{\mathbf{k}}$, $\zeta_{\mathbf{k}}$ and $A^{\pm}$, discussed in Eq.~\ref{Eq_Hk_1} and Eq.~\ref{Eq_A_pm}, we evaluate the integrals $J_0^{\pm}$, $I_0^{\pm}$, and $K_0^{\pm}$ numerically in the limit $\omega\rightarrow 0$, to obtain $G_0(\mathbf{0},\omega)$.  We note that $J_0^+ = J_0^-$, because $F_{\mathbf{k}}$ is an odd function of $\mathbf{k}$. Therefore, for a single magnetic impurity with Ising SOC, the problem is identical to the case of a magnetic impurity in an $s$-wave superconductor without SOC~\cite{Brydon:2015}. Due to the localized $\delta$-function nature of the magnetic impurity potential, the integrals involving the spin-triplet pairing terms in the Green’s function vanish. 

Once $G_0(\mathbf{r}=\mathbf{0},\omega)$ is obtained numerically, we can then solve Eq.~\ref{Eq_bdg_1} for the impurity bound state $\Psi(\mathbf{r}=0)$ (which is the YSR state). In our analysis, we limit ourselves only upto $\omega$- linear terms, since we are interested in solutions which lie close to the center of the superconducting gap.  Denoting the matrix $G_0(\mathbf{r}=\mathbf{0},\omega) J\mathbf{S}\tau_0\hat{\sigma}$ in Eq.~\ref{Eq_bdg_1} as $\omega\mathcal{L}+\mathcal{M}$ (the matrices $\mathcal{L}$ and $\mathcal{M}$ are determined numerically and are now independent of $\mathbf{k}$ and $\omega$, and only depend on the material parameters), the subgap spectrum for the YSR state is then given by~\cite{Brydon:2015}
\begin{eqnarray}
-(\mathcal{L})^{-1} (\mathbb{I}+\mathcal{M}) \Psi(\mathbf{0}) = \omega\Psi(\mathbf{0})
\label{Eq_Heff_1}
\end{eqnarray}
Eq.~\ref{Eq_Heff_1} can be directly solved for $\Psi(\mathbf{0})$ and the energy spectrum $\omega$. Also, Eq.~\ref{Eq_Heff_1} can be solved to obtain the critical exchange strength $J_c$, such that when $J=J_c$, the YSR state spectrum admits a solution at exactly zero energy ($\omega=0$). In Figure~\ref{Fig_Shiba_spec} we have plotted the energy levels as a function of the impurity strength $J$, obtained by numerically solving Eq.~\ref{Eq_Heff_1}. The figure highlights the existence of a critical impurity strength $J_c$, where two mid-gap energy levels cross each other at $\omega=0$. The inset of Fig.~\ref{Fig_Shiba_spec} also shows the two energy levels which are away from mid-gap region. This mid-gap zero energy YSR bound state emerging due to
a single localized magnetic impurity located on a $s$-wave superconductor
gives rise to a ZBP in the
local density of states measurement as discussed in Sec. III. 

\begin{figure}
\includegraphics[scale=.22]{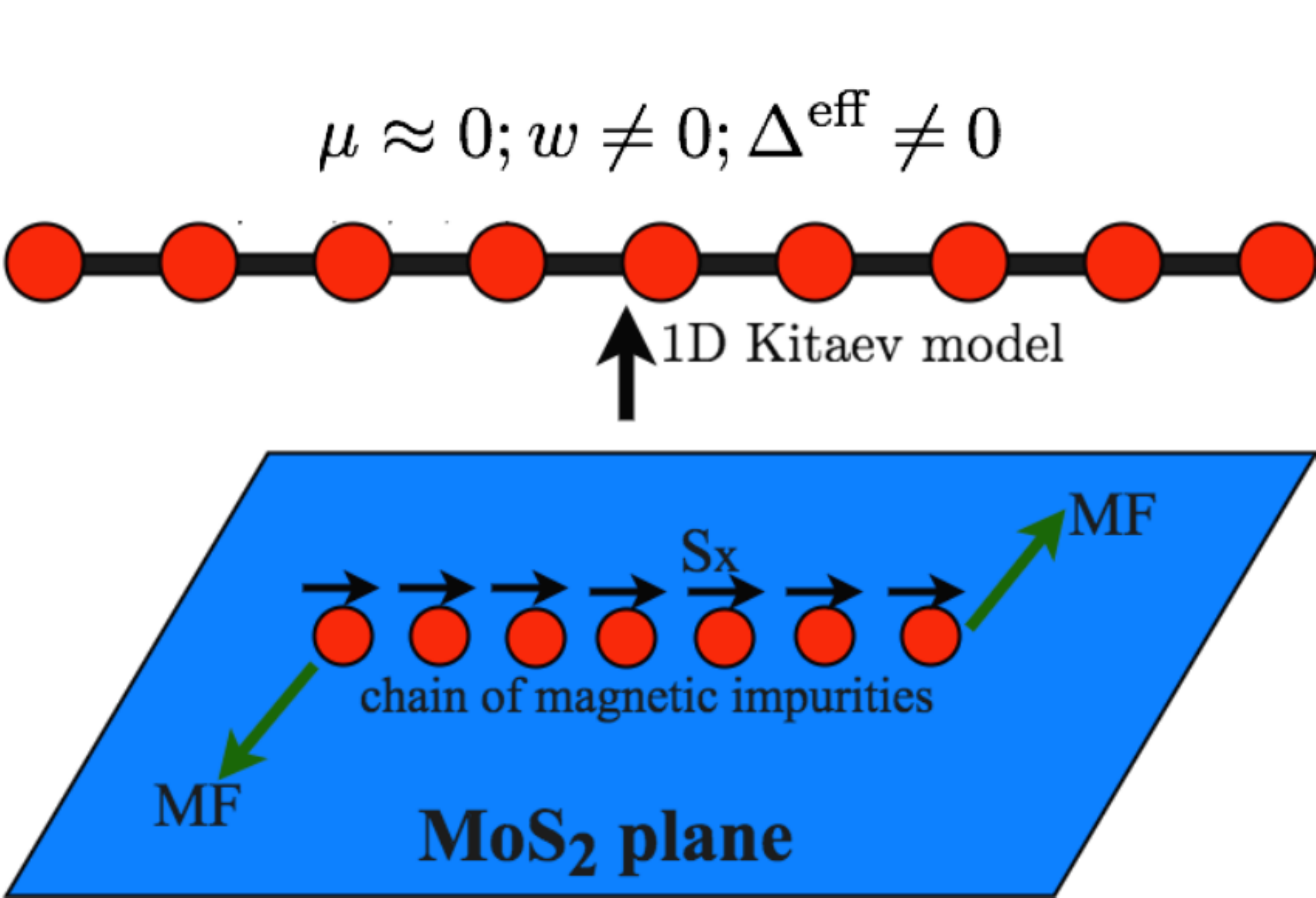}
\includegraphics[scale=.21]{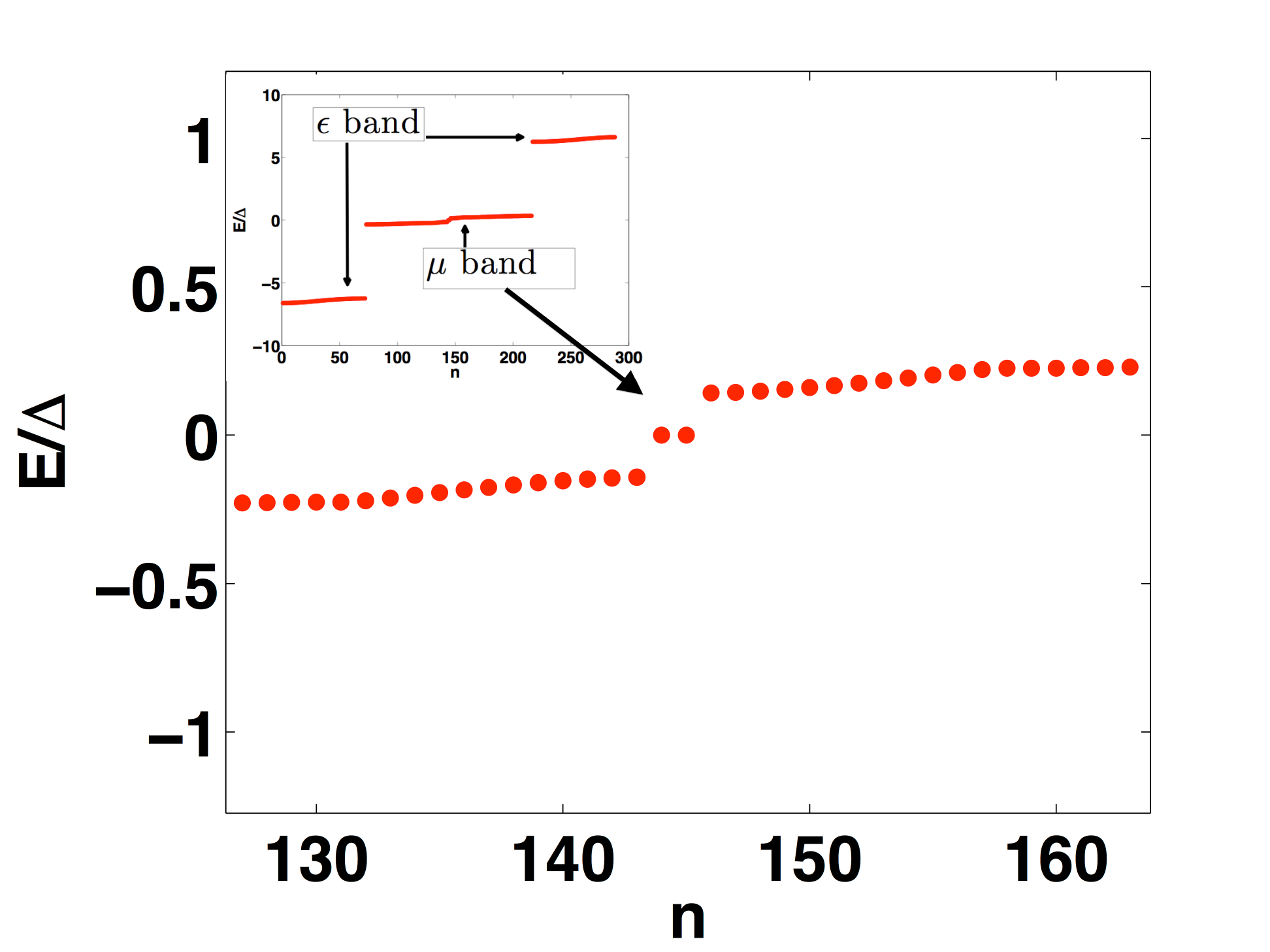}
\includegraphics[scale=.21]{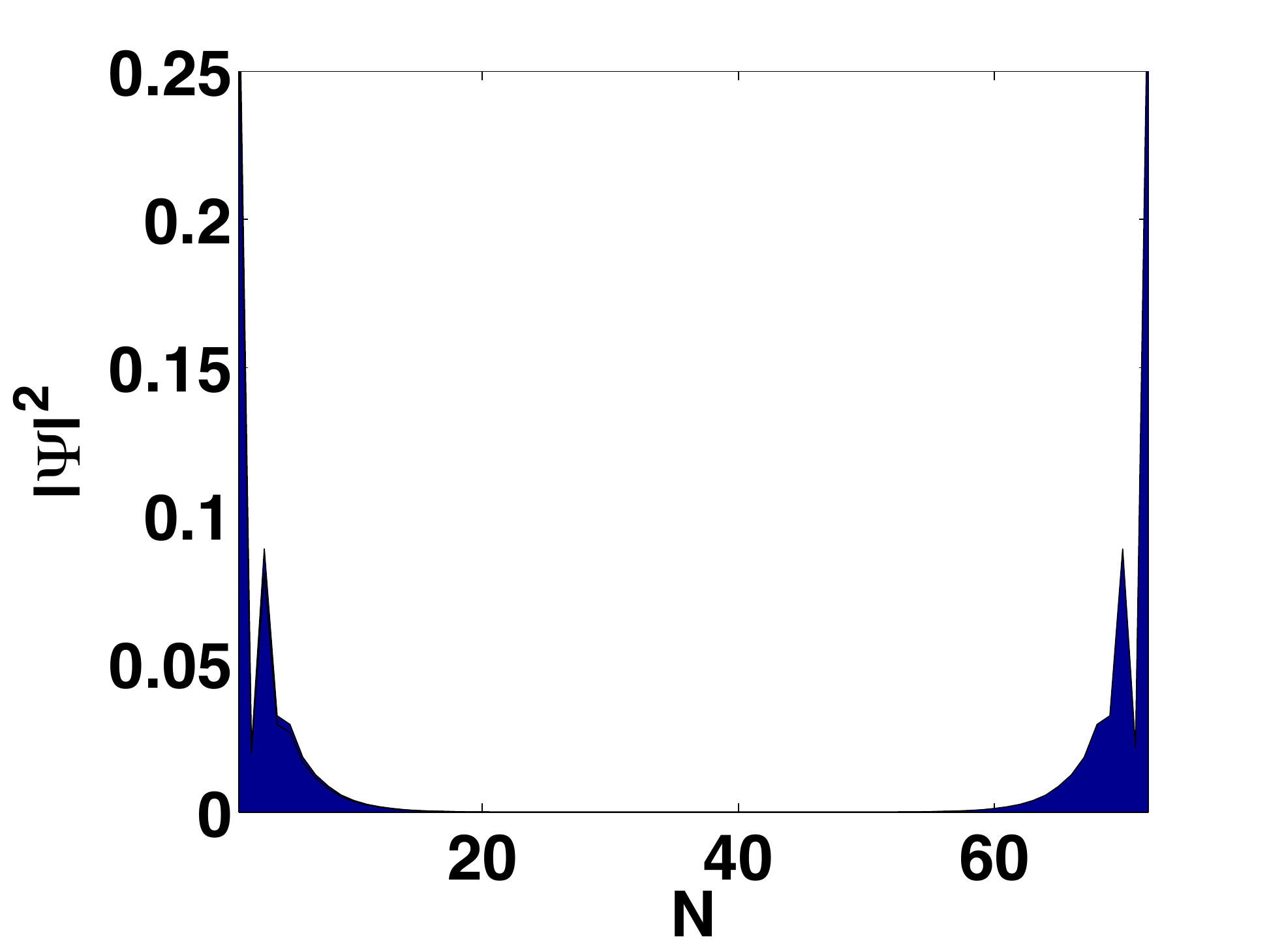}
\caption{(color online). \textit{Top panel:} Schematic diagram of a ferromagnetic impurity chain embedded on superconducting MoS$_2$ surface, with spins pointing parallel to the plane. The effective tight-binding model can be mapped onto the 1D Kitaev model resulting in localized Majorana modes at the two ends. \textit{Bottom panel:} Energy spectrum vs. eigenvalue index for the real space BdG equation (Eq.~\ref{Eq_Hbdg_3} written in the basis of individual YSR states), illustrating an induced superconducting gap with two zero energy Majorana modes. On the right panel, the corresponding wavefunctions are plotted showing localization near the edges. The number of impurity sites was taken to be $N=72$, with impurity spacing $d=4a$, where $a$ is the lattice spacing. Further, the spins of the impurities were aligned along the $x$ direction, and only nearest neighbor coupling between the impurity modes was assumed. The value of critical exchange strength chosen was $J=0.3925$. }
\label{Figure_Majorana_1}
\end{figure}

In order to discuss topological superconductivity, we will now consider a ferromagnetic chain of impurities embedded on superconducting MoS$_2$ substrate. Assuming the chain runs along the $x$ direction, the impurity Hamiltonian becomes
\begin{eqnarray}
H_{\text{imp}} = \sum\limits_{i}{[\Psi^\dagger(x_i) (-J\mathbf{S}\tau_0\hat{\sigma})\Psi(x_i)]}
\label{Eq_Himp_2}
\end{eqnarray}
The BdG equation (Eq.~\ref{Eq_bdg_1}, ~\ref{Eq_Heff_1}) can now be generalized to~\cite{Pientka:2013, Brydon:2015}
\begin{eqnarray}
&(\mathbb{I} + \omega\mathcal{L}+\mathcal{M})\Psi(x_i) =
&-\sum\limits_{j\neq i}{I_{ij} (J\mathbf{S}\tau_0\hat{\sigma})\Psi(x_j)},
\label{Eq_Hbdg_2}
\end{eqnarray}
where $I_{ij}$ is the correlator $I_{ij}=G_0(x_i-x_j,\omega)$ which generates an effective coupling between the individual YSR states at impurity site $x_i$ and $x_j$. We will work in the regime where the decoupled impurity states occur at energies close to $\omega=0$. Such a condition is guaranteed to occur when the exchange strength is tuned near the critical impurity exchange strength $J_c$, as already highlighted in Figure~\ref{Fig_Shiba_spec}. Now when the coupling $I_{ij}$ is turned on, it hybridizes the YSR states to drift away from $\omega=0$ to form a YSR band near the mid-gap. Analogous to the single impurity problem, we need to evaluate the following integrals in order to compute $I_{ij}$ in the limit $\omega\rightarrow 0$~\cite{Pientka:2013, Brydon:2015}
\begin{eqnarray}
&J_1^{\pm} = \int{\frac{[d^2\mathbf{k}]e^{ik_x(x_j-x_i)}}{A^{\pm}}}\nonumber\\
&I_1^{\pm} = \int{\frac{[d^2\mathbf{k}]e^{ik_x(x_j-x_i)}\zeta_{\mathbf{k}}}{A^{\pm}}}\nonumber\\
&K_1^{\pm} = \int{\frac{[d^2\mathbf{k}]e^{ik_x(x_j-x_i)}F_{\mathbf{k}}}{A^{\pm}}}
\end{eqnarray}
The coupling $I_{ij} = G_0(x_i-x_j,\omega)$ is then given by
\begin{eqnarray}
I_{ij}=\left( \begin{array}{cccc}
  I_1^+ + K_1^+ &0 & \Delta J_1^+ & 0 \\
0 &   I_1^- - K_1^-  & 0 & \Delta J_1^-\\
\Delta J_1^+ & 0 &  - I_1^+ - K_1^+  & 0\\
0 &\Delta J_1^-  & 0 &  - I_1^- + K_1^-
\end{array} \right) \nonumber\\
\end{eqnarray}
In contrast to the single magnetic impurity problem discussed earlier, the presence of Ising SOC significantly affects the BdG equations. When the Ising SOC parameter $\alpha\neq 0$, $J_1^+\neq J_1^-$, implying a non-zero superconducting triplet correlation in the Green's function $G_0(x_i-x_j)$. This feature gives rise to a non-zero effective $p$-wave superconducting component as required for topological superconductivity. The couplings $I_{ij}$ can been computed using numerical integration over the 2D Brillouin zone. The BdG equation (Eq.~\ref{Eq_Hbdg_2}) can then be rewritten in the following form after evaluating the couplings in the limit $\omega\rightarrow 0$.
\begin{eqnarray}
&-&(\mathcal{L})^{-1} (\mathbb{I}+\mathcal{M})\Psi(x_i) \nonumber \\
&+& \sum\limits_{j\neq i} (-\mathcal{L})^{-1} I_{ij} (\omega\rightarrow 0) (J\mathbf{S}\tau_0\hat{\sigma})\Psi(x_j)
 = \omega\Psi(x_i)
 \label{Eq_Hbdg_3}
\end{eqnarray}
This equation can now be projected on to the basis of individual YSR states, to obtain an effective tight-binding Hamiltonian which can be mapped on to a 1D Kitaev model for a topological superconductor with long range couplings between various impurity sites~\cite{Pientka:2013, Brydon:2015, Kitaev:2001}.
\begin{eqnarray}
H^{\text{eff}} = \sum\limits_i\sum\limits_{j\neq i} {h^0_i + h^n_{ij} + \Delta^{\text{eff}}_{ij}}
\label{Eq_tight_binding}
\end{eqnarray}
The term $h^0_i = \epsilon\gamma_z\beta^+ + \mu\gamma_z\beta^-$, where $\pm\epsilon$ and $\pm\mu\approx 0$ are now the energy levels of the uncoupled YSR states (as illustrated in Figure~\ref{Fig_Shiba_spec}, not to be confused with $\epsilon$ and $\mu$ in Sec II, where they stand for the valley index and chemical potential respectively). The Pauli matrix $\beta$ now acts on the inter $\epsilon$-$\mu$ energy space, and $\gamma$ acts on the intra $\epsilon$-$\mu$ energy space. The energy levels $\pm \mu$ lie close to the midgap, while the levels $\pm \epsilon$ are away from the midgap. Therefore these form the $\epsilon$ band (or the $\beta^+$ band) and the mid-gap $\mu$ band (or the $\beta^-$ band, where $\mu\approx 0$), in the absence of any couplings. The term $h^n_{ij}$ is the effective hopping integral between sites $i$ and $j$, and $\Delta^{\text{eff}}_{ij}$ is the induced  effective superconducting parameter. We evaluate these terms and retain couplings only upto nearest neighbor for our calculations. We find $h^n_{i,i+1}$ to be of the form  $h^n_{i,i+1} = w\beta^- + v\beta^+$, where $w$ and $v$ have been evaluated numerically. Furthermore, $\Delta^{\text{eff}}_{i,i+1}$ is evaluated to be of the form $\Delta^{\text{eff}}_{i,i+1} = \delta^+\beta^+ + \delta^-\beta^-$, which is the $p$-wave superconducting order parameter, also inspected to be non-vanishing if the Ising SOC $\alpha\neq 0$, and spins of the magnetic impurities lie parallel to the MoS$_2$ plane. Physically, this is expected because a Zeeman type field (which is generated by the impurity spins in present case) parallel to the spin-orbit field will not create a quasiparticle gap in the spectrum and thus will not induce topological superconductivity akin to the 1D semiconductor Majorana wire platform.

On numerical diagonalization of the real space BdG equation (Eq.~\ref{Eq_Hbdg_3} projected onto individual YSR states), the midgap YSR states ($\beta^-$ band) hybridize away from $\omega\approx 0$, however two protected Majorana edge modes, and an induced superconducting gap appear in the YSR band, when the impurity spins are aligned along the $x$ direction. Figure~\ref{Figure_Majorana_1} shows the energy spectrum, illustrating an induced superconducting gap with two zero energy Majorana modes, for a 1D chain of 72 sites. The corresponding wave-functions show localization near the edges of the chain. Though, for our calculations we assumed only nearest neighbor interaction between the sites, we have checked that the emergent topological superconductivity remains intact by including longer range hopping and pairing terms in the effective tight-binding Hamiltonian.
\section{Conclusions}
TMDs are materials with 2D honeycomb lattice similar to graphene, but have broken in-plane mirror symmetry, resulting in a special type of intrinsic spin-orbit coupling, called Ising SOC. Ising SOC acts as an effective Zeeman field which strongly polarizes the electron spins perpendicular to the 2D plane. Interestingly, the spin polarization is not constant in momentum space, but rather changes sign across the $\Gamma$ point, which gives rise to a very high in-plane critical magnetic field in superconducting TMDs~\cite{{Lu:2015},{Saito:2015},{Xi2:2015}}. 
In this work we showed that the magnetic field response of STM zero bias peaks from magnetic adatoms in Ising superconductors is strongly anisotropic (with critical Zeeman fields  for ZBP splitting applied parallel and perpendicular to Ising SOC being in the ratio $\sim 1:32$). This behavior of YSR states, a direct consequence of Ising SOC, is of immediate experimental interest.
Furthermore, this response also correlates well with the anomalously large anisotropy in upper critical fields between directions perpendicular and parallel to the 2D plane as revealed in recent experiments.

Further, we show the emergence of a topological superconducting phase in the impurity YSR band for a dilute concentration of magnetic impurities arranged in a chain-like configuration with moments parallel to the plane of the superconductor. In the topological superconducting phase, zero energy Majorana fermions appear at the ends of impurity chain and can be accessed by scanning tunneling microscopy experiments as in the recent experiments on chains of Iron impurities on spin-orbit coupled Pb superconductor~\cite{Perge2:2014}. In contrast to the case of the Pb superconductor we find that in order to support a topological superconducting phase the magnetic moments of the impurities embedded in Ising superconductors need to be parallel (or have a parallel component) to the plane of the superconductor. This is a direct consequence of Ising SOC which consists of an effective $k$-dependent Zeeman field perpendicular to the 2D plane. That the direction of the SOC should be transverse to the direction of the Zeeman field (which in the present case is given by the magnetic moments) for the existence of the topological superconducting phase is also true in the models of topological superconducting phase in spin-orbit coupled superconductor-semiconductor heterostructures~\cite{Sau1:2010,Sau2:2010,Oreg:2010,Lutchyn:2010}. 

In this paper we considered the limit of dilute concentration of impurities embedded in the Ising superconductor. In the complementary band (or `wire') limit, the impurity chain realizes a topological superconductor (in BDI class) by proximity effect \cite{Hui:2015,Dumitrescu:2015}, similar to the case of a half-metal \cite{Zhou:2015}.
We thus establish Ising superconductors with magnetic adatoms with moments parallel to the host superconductor as a robust platform for topological superconductivity and Majorana excitations which can be probed in STM experiments.

\textit{Acknowledgement.}
The authors acknowledge discussions with J. D. Sau and support from AFOSR (FA9550-13-1-0045).

\textit{Note added:}
Recently another manuscript (Ref~\onlinecite{ZhangAji}) appeared which draws similar conclusions. In Ref.~\onlinecite{ZhangAji} it is also concluded that TMDs can support a topological superconducting state in the YSR chain, as long as the magnetic moments have a finite in-plane component. This result is consistent with our result, that adatom moments must be parallel (or have a parallel component) to the 2D plane. This is a crucial point in both works and the important difference from a Rashba superconductor. Additionally, we also study the magnetic field response of STM zero bias peaks from magnetic
adatoms in Ising superconductors. 

\end{document}